\pdfoutput=1
\documentclass[11pt]{article}

\usepackage[a4paper,margin=1in]{geometry}
\usepackage{lmodern}      
\usepackage[T1]{fontenc}
\usepackage[utf8]{inputenc}
\usepackage{microtype}
\usepackage{setspace}     
\usepackage{authblk}

\usepackage{amsmath,amssymb,amsfonts,bm} 
\usepackage{graphicx}
\usepackage{booktabs}
\usepackage{adjustbox}
\usepackage{algorithm,algpseudocode}
\usepackage[numbers,sort&compress]{natbib} 
\usepackage{hyperref}
\usepackage{cleveref} 
\usepackage{subcaption}
\usepackage{enumitem}
\usepackage{threeparttable}
\usepackage{pdflscape}

\title{Quantifying Fire Risk Index in Chemical Industry Using Statistical Modeling Procedure}

\author[1]{Hyewon Jung}
\author[1]{Seungil Ahn}
\author[1]{Seungho Choi}
\author[2,*]{Yeseul Jeon}

\affil[2]{Department of Epidemiology \& Biostatistics, University of California, San Francisco; 
Department of Statistics, Texas A\&M University.}
\affil[1]{Korea Fire Protection Association, Seoul, Korea.}

\affil[*]{Corresponding author: \texttt{jeons9677@tamu.edu}}

\date{} 

\begin{document}
\maketitle

\begin{abstract}
Fire incident reports contain detailed textual narratives that capture causal factors often overlooked in structured records, while financial damage amounts provide measurable outcomes of these events. Integrating these two sources of information is essential for uncovering interpretable links between descriptive causes and their economic consequences. To this end, we develop a data-driven framework that constructs a composite Risk Index, enabling systematic quantification of how specific keywords relate to property damage amounts. This index facilitates both the identification of high-impact terms and the aggregation of risks across semantically related clusters, thereby offering a principled measure of fire-related financial risk. Using more than a decade of Korean fire investigation reports on the chemical industry classified as Special Buildings (2013–2024), we employ topic modeling and network-based embedding to estimate semantic similarities from interactions among words, and subsequently apply Lasso regression to quantify their associations with property damage amounts, thereby estimate fire risk index. This approach enables us to assess fire risk not only at the level of individual terms but also within their broader textual context, where highly interactive related words provide insights into collective patterns of hazard representation and their potential impact on expected losses. The analysis highlights several domains of risk, including hazardous chemical leakage, unsafe storage practices, equipment and facility malfunctions, and environmentally induced ignition. The results demonstrate that text-derived indices provide interpretable and practically relevant insights, bridging unstructured narratives with structured loss information and offering a basis for evidence-based fire risk assessment and management. The derived Risk Index provides practical reference data for both safety management and insurance underwriting by enabling the prioritization of preventive measures within industrial sites and offering quantitative guidance for assessing facility-specific risk levels in insurance decisions. An R implementation of the proposed framework is openly available for public use.
\end{abstract}


\section{Introduction}
Fire accidents pose serious threats to industrial safety, human life, and the economy. In particular, chemical facilities are highly vulnerable because of their dependence on flammable and reactive materials, where minor ignition can escalate into catastrophic events. Understanding and quantifying such risks are therefore essential for developing effective preventive measures and for guiding insurance and safety management policies. While structured fire statistics provide valuable information about fire frequency and damages, they often fail to capture the complex causal mechanisms recorded in textual fire investigation narratives. These reports, written by on-site officers, contain unstructured yet highly informative descriptions that reflect contextual factors such as process operations, equipment malfunctions, material leakage, and human or environmental interactions. Leveraging such unstructured narratives through text mining offers a promising pathway for discovering latent risk patterns that cannot be detected from aggregated statistics alone.

Previous research has employed a variety of approaches to analyze fire-related risks. Text-based studies have demonstrated the potential of mining unstructured documents. For instance, \citet{kim2020text} applied topic modeling techniques to accident verdicts from ship fire cases to identify ignition sources, flammable materials, and negligence-related causes, while \citet{tirunagari2015data} investigated maritime accident investigation reports using text mining to extract causal relations among contributing factors. Although these studies represented important early steps in applying text analytics to fire accidents, they primarily relied on word-frequency clustering and did not capture the correlated structure among words that reflects the interdependent nature of fire risk factors. More recently, \citet{liu2023forestfire} analyzed scientific literature to trace evolving research trends in forest fire studies, and \citet{zhao2024wui} constructed a comprehensive database of wildland–urban interface fires. Similar text mining frameworks have been used to identify risk factors in ship fires \citep{zhang2025shipfire}, heritage building fires \citep{wang2023heritagevar}, and industrial settings such as mining and chemical enterprises \citep{wang2022chemicalbn, li2022coalrisk, liu2022hotwork, yang2023coalchemical, chen2024triangle}. Narrative-based perspectives have also been explored: for example, \citet{russo2024wildfire} analyzed wildfire narratives to identify multiple social storylines concerning causes, consequences, and potential solutions, illustrating how unstructured narratives can provide a contextualized understanding of fire events and their socio-environmental implications.

Structured and indicator-based approaches have likewise received considerable attention. \citet{ma2025occurrence} conducted a data-driven analysis of over one million building fire reports, integrating structured incident records with socioeconomic and structural variables to assess the effects of detection systems and extinguishing devices on fire spread and injury risk. Similarly, \citet{zhang2025developing} developed an indicator system for urban fire risk assessment that emphasizes meteorological and building characteristics. Other domain-specific efforts include quantitative frameworks for Value-at-Risk of heritage structures \citep{wang2023heritagevar}, climate-related hazard assessment \citep{dapeng2025wui}, and chemical industry safety evaluation \citep{kudryavtsev2025chemical}. However, most of these structured approaches exclude unstructured textual data, thereby omitting rich qualitative details that often contain the causal reasoning behind fire events.

Parallel progress in natural language processing and machine learning has broadened the methodological basis for text-based risk analytics. Transformer-based models such as Sentence-BERT have improved short-text clustering by modeling contextual similarity between sentences \citep{kim2024sbert, park2024sberttransfer}, while hybrid topic–autoencoder architectures enhance detection of latent thematic structures \citep{choi2024topicauto, lee2024autoencoder}. Graph-based and network-analytic frameworks have also been proposed to represent relationships among semantic entities, enabling discovery of interaction mechanisms in multilingual or domain-specific corpora \citep{han2023graph, jeong2025network, zhou2024tfsc}. These methodological advances have improved text clustering, anomaly detection, and classification across diverse fields, including risk disclosure analysis \citep{risk2023financial}, software fault diagnosis \citep{elsevier2024software}, and insurance claim interpretation \citep{rahman2024insurance}. Yet, existing fire-related applications remain largely descriptive, focusing on frequent keywords or co-occurrence patterns rather than connecting textual semantics to measurable outcomes such as financial or property loss.

Taken together, the literature reveals a persistent methodological and practical gap. Existing fire text-mining studies seldom model the dependent structure among risk-related terms that jointly describe operational and environmental conditions within accident narratives. Likewise, few studies link textual features to structured indicators such as damage costs or insurance data. Consequently, current approaches cannot quantify how textual expressions of risk translate into tangible loss or prioritize critical risk factors by their financial relevance. Addressing this gap, the present study introduces a unified framework that systematically analyzes fire investigation narratives, estimates semantic dependencies among risk-related terms, and integrates these text-derived features with estimated damages data to construct a composite, loss-aware fire Risk Index. This approach bridges unstructured and structured domains, offering a scalable and interpretable foundation for quantitative fire risk analysis and decision-making.

This framework is expected to be integrated into industrial safety management and insurance risk-assessment systems. The derived Risk Index provides quantitative reference data that enable facility managers to prioritize preventive measures according to site-specific hazards extracted from textual records. It also assists insurers in evaluating facility-level risks that cannot be captured by conventional loss-ratio statistics. By linking narrative investigation data with measurable financial outcomes, the framework facilitates data-driven decision making for both workplace safety management and insurance underwriting.

\paragraph{\textbf{Contributions}}  
This study makes the following contributions:  
\begin{itemize}  
    \item \textbf{First large-scale analysis of Korean fire investigation narratives.} To the best of our knowledge, this is the first systematic attempt to analyze over a decade (2013–2024) of textual records on the chemical industry classified as Special Buildings, written by fire officers. These narratives provide direct accounts of ignition sources and causal conditions that have not been previously examined in quantitative fire safety research. 
    \item \textbf{Integration of textual evidence with economic loss indicators.} By linking unstructured narratives with structured data on fire property damage, we move beyond frequency-based measures of risk. This integration enables the identification of risk factors that matter not only for their occurrence but also for their \emph{economic impact}, thereby offering a more practical and policy-relevant assessment of fire risk.  

    \item \textbf{Development of a statistically grounded risk index.} We combine established statistical approaches~\citep{jeon2025network} with a Lasso regression framework to construct a novel composite index. This allows us to capture meaningful dependency structures among words and quantify their contribution to observed loss outcomes. The resulting index provides an interpretable and efficient tool to quickly identify critical fire-related terms that elevate risk.  
\end{itemize}

The remainder of this paper is organized as follows.
Section 2 introduces the dataset and outlines the analytical procedures, including latent topic estimation via the Biterm Topic Model, topic clustering through a Latent Space Item Response Model, and the construction of a risk index factor using Lasso regression.
Section 3 presents the results, covering topic characterization through words, thematic aggregation of topics, and evaluation of the proposed risk index factor.
Section 4 discusses the implications of the findings and concludes the study.


\section{Materials and Methods}

\begin{figure}[ht]
\centering\includegraphics[scale=0.20]{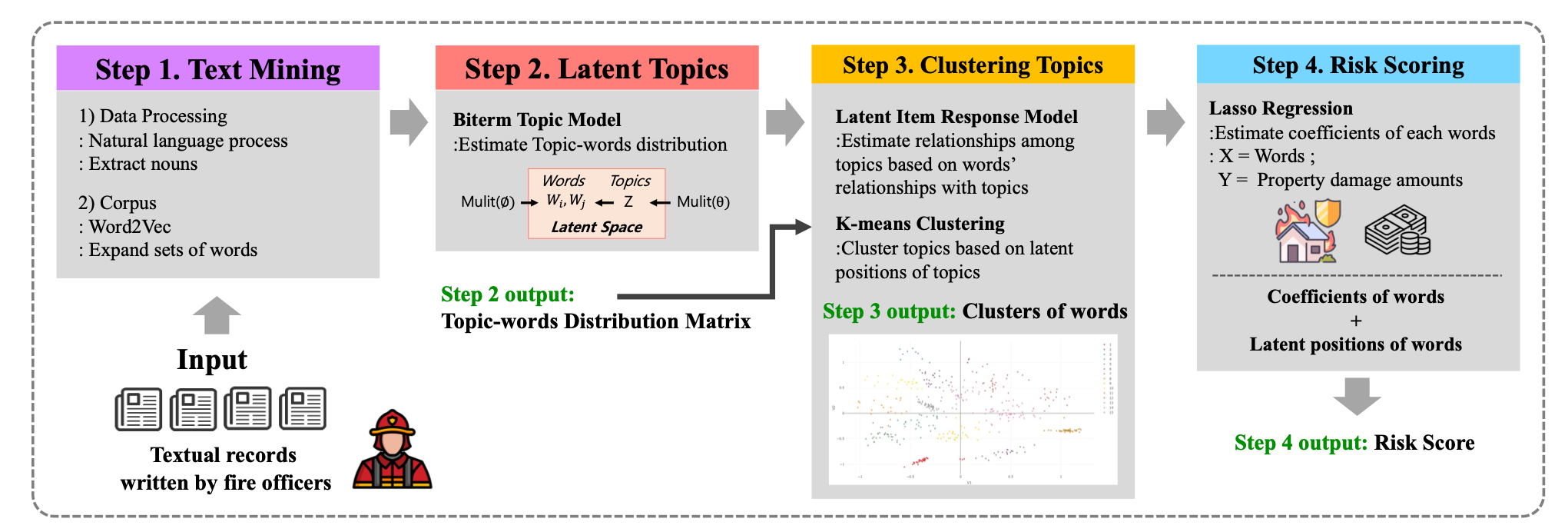}
\caption{Analytical framework for constructing a fire risk index from unstructured investigation texts.
The process begins with text mining to preprocess investigation records and extract key nouns (Step 1). A Biterm Topic Model is then applied to estimate topic–word distributions (Step 2). Next, a latent item response model is used to infer topic–word interactions and to estimate latent positions of words, which are subsequently clustered to capture semantic groupings (Step 3). Finally, Lasso regression estimates the coefficients of words using property loss data as the outcome, yielding a composite risk index that integrates words’ semantic closeness with structured information derived from loss amounts (Step 4).}
\label{fig:overall}
\end{figure}

Figure~\ref{fig:overall} presents the proposed analytical framework for constructing a fire risk index from unstructured fire investigation texts. In the first step (Step 1), natural language processing techniques are applied to preprocess the textual records and extract nouns, with a particular emphasis on building a domain-specific lexicon related to the chemical industry. This enables the identification and expansion of keywords that are directly relevant to chemical processes and accident scenarios, providing a structured corpus for further analysis. 

In the second step (Step 2), a Biterm Topic Model~\citep{yan2013biterm} is employed to classify documents into latent topics and to estimate the distribution of words within each topic. This step not only offers a compact summary of large-scale documents but also transforms unstructured text into a topic–words distribution matrix that can be further exploited. Using these topic–word distributions as input for the subsequent latent space model is advantageous, as it embeds words into a representation that reflects their co-occurrence patterns, thereby facilitating the estimation of meaningful word–word interactions that would not be apparent from raw text alone.

In the third step (Step 3), the latent item response model~\citep{Jeon:2021} is applied to infer the positions of words in a continuous latent space, which captures their semantic relationships. The estimated distances between words can be interpreted as measures of semantic similarity: for example, if two words are placed close to each other in the latent space, they are more likely to represent semantically related concepts. Leveraging these latent positions, words are clustered into semantically coherent groups, thereby enabling the construction of interpretable clusters of risk-related vocabulary. Importantly, the estimation of interword interactions accounts for both the frequency and contextual sparsity of co-occurrences. Thus, words are not measured as strongly associated merely because they occasionally appear together; rather, their latent positions are inferred based on consistent and dense interaction patterns with other words across the corpus.

In the final step (Step 4), structured data on property damage amounts are incorporated into the analysis. Specifically, Lasso regression is used to estimate coefficients linking each word to the magnitude of property loss. Words with higher coefficients can thus be interpreted as risk factors associated with greater expected damages. Beyond word-level inference, this integration allows for cluster-level analysis: by examining the aggregated coefficients of words within each cluster, we can identify which semantic groupings correspond to high-risk factors in terms of potential financial losses. Taken together, this framework not only provides a systematic way to quantify fire risk from unstructured narratives but also bridges semantic information extracted from text with structured loss data to yield interpretable and practically meaningful risk indices.

\subsection{Data}
This analysis focuses on fire accidents in Special Buildings in Korea that fall under the chemical industry (excluding plastic production) within the category of Manufacturing uses. In Korea, Special Buildings are legally designated facilities identified as high fire-risk based on their intended use, scale, and other criteria. These facilities are overseen by the Korea Fire Protection Association (KFPA) for fire prevention purposes. As of the end of 2024, there were 54,517 such buildings nationwide, and the total number changes annually.

The dataset used in this study consists of records from post-incident investigations of fire accidents in these chemical industry Special Buildings between 2013 and 2024. As shown in Table~\ref{tab:narrative_examples}, the fire investigation records are unstructured text documents describing the fire circumstances, such as the cause, involved equipment, and ignited materials. All investigation texts are written in Korean. In addition, the dataset includes property damage estimates verified by local fire departments in accordance with the Fire Damage Assessment Manual of the National Fire Agency of Korea, covering both movable and immovable property losses such as buildings, facilities, machinery, and inventory assets.

\begin{table}[H]
\centering
\scriptsize
\caption{Ten randomly selected representative excerpts of fire investigation records}
\label{tab:narrative_examples}
\begin{adjustbox}{max width=\linewidth}
\begin{tabular}{|c|p{13cm}|}
\hline
\textbf{ID} & \textbf{fire investigation records} \\
\hline
1 & A fire occurred in the vulcanization process of Plant 2. Flames and smoke were first observed under the pit of equipment No. 3115 on the 3100 line. The area showed severe damage and electrical abnormalities, and CCTV confirmed early smoke from the same location. The fire is presumed to have originated beneath the pit of No. 3115. \\
\hline
2 & In the production building of a factory, sunflower oil was being refined by mixing oil, hexane, and activated carbon with a stirrer. Hexane vapors leaked and ignited from an unidentified ignition source, causing an explosion with fire. \\
\hline
3 & During repair on the duct connected to a dust collector, grinder sparks ignited sponge soundproofing material on the wall. Scene investigation indicated initial ignition at the wall near the duct; burn patterns showed vertical spread to the second floor. The cause was determined as worker negligence. \\
\hline
4 & Prior to an explosion and fire, an emergency call reported injuries due to steam leakage. Investigation found that unreacted photoinitiator was normally returned to a 5 ton reactor, but aluminum chloride catalyst was mistakenly added through a manhole, causing ejection. While workers evacuated, solvent vapors accumulated and later ignited from an unknown source, causing an explosion and fire. \\
\hline
5 & A witness heard a beep and saw vapor leaking from a reactor lid. After donning a respirator, larger vapor was seen from the lid, condenser, and receiver tank area; droplets and chemicals fell from the reactor bottom. A brief initial explosion occurred, followed by a major explosion seconds later. \\
\hline
6 & In a pharmaceutical production plant, an explosion and fire occurred during blending and drying of raw materials in the third floor mixing and granulation room. The fire spread and damaged a large portion of equipment and facilities. \\
\hline
7 & A three story chemical plant producing BP F 95 experienced a fire presumed to have originated inside a centrifugal dehydrator while dehydrating BP F 95 mixed with toluene. The fire partially spread up to the third floor; no injuries were reported. \\
\hline
8 & An explosion occurred due to a reaction between reactor contents and unidentified static electricity in a synthesis plant. The fire spread to nearby organic solvents, resulting in total destruction of the synthesis building. \\
\hline
9 & In the drying room of Building D, an operator was loading wet powder cosmetic raw material into a dryer tray. Acetone vapors reached an explosive range and presumably ignited within the operator’s working radius, spreading to nearby combustibles. \\
\hline
10 & No other heat source was identified at the origin except sparks from welding work. Burn patterns indicated rapid spread from stored flammable material (styrofoam). Based on witness statements and proximity to the welding site, welding sparks were presumed to have ignited the styrofoam. \\
\hline
\end{tabular}
\end{adjustbox}
\end{table}

By leveraging more than a decade of investigation reports (2013–2024), the analysis captures long-term and stable patterns of fire incidents within a relatively homogeneous industrial environment. This extensive temporal coverage mitigates sample-specific variability and enhances the generalizability of the derived semantic structures.

A total of 8,190 fire investigation reports were analyzed, from which 449 representative keywords were selected based on three filtering criteria: excluding 763 non-informative terms (e.g., numbers, symbols, general words, or names), consolidating 315 synonymous words into 165 representative entries, and retaining only those appearing at least twice across all reports.

\subsection{Latent Topic Estimation via the Biterm Topic Model}

We employ the Biterm Topic Model (BTM) to uncover latent semantic structures within textual statements written by firefighters during post-incident investigations of fire causes. As a preprocessing step, morphological analysis is conducted to decompose words into their base morphemes. From these, we extract nouns in their canonical form, which constitute the initial corpus. To enrich this corpus, we expand the vocabulary set using the {\tt word2vec} algorithm, which maps words into a Euclidean latent space according to semantic similarity. By identifying words in close proximity within this embedding space, we obtain an augmented corpus that better captures the semantic landscape of the texts. 

The ultimate objective is to identify a collection of latent topics that summarize the semantic content of the documents. Each topic is represented as a distribution over words, while each document is modeled as a mixture of these latent topics. Since those reports are typically consist of fewer than 200 words and can thus be regarded as short texts, the BTM is particularly suitable for this setting. The model relies on extracting biterms, i.e., unordered word pairs within documents, which serve as the input to the topic model.

The BTM is founded on several key assumptions: 
\begin{itemize}
    \item Each pair of words (biterm) is assumed to arise from an underlying latent topic. 
    \item Topics themselves represent semantically coherent clusters of words. 
    \item Word co-occurrence patterns within the corpus can therefore be explained through mixtures of such latent topics. 
\end{itemize}

\noindent
Formally, the likelihood of BTM is determined by the topic distribution and the topic–word distributions. Two sets of parameters must therefore be estimated: the topic proportion vector $\boldsymbol{\theta}$ and the topic–word distributions $\boldsymbol{\phi}_{z}$. The prior for each $\boldsymbol{\phi}_{z}$ is specified as a Dirichlet distribution with hyperparameter $\beta$, while the prior for $\boldsymbol{\theta}$ follows a Dirichlet distribution with hyperparameter $\alpha$. A latent topic assignment variable $z$ is drawn from a Multinomial distribution with parameter $\boldsymbol{\theta}$, and conditional on $z$, each word is generated from $\boldsymbol{\phi}_{z}$. Hence, the parameters of interest are $\boldsymbol{\theta}$, $\boldsymbol{\phi}_{z}$, and $z$.

The generative process of BTM can be summarized as: 
\begin{enumerate}[leftmargin=*,labelindent=14pt, label=Step \arabic*]
    \item Draw a topic distribution $\boldsymbol{\theta} \sim \text{Dirichlet}(\alpha)$. 
    \item For each biterm $b \in B$, assign a latent topic $z \sim \text{Multinomial}(\boldsymbol{\theta})$. 
    \item For each topic $z$, draw a topic–word distribution $\boldsymbol{\phi}_{z} \sim \text{Dirichlet}(\beta)$. 
    \item Generate the two words $w_i, w_j \sim \text{Multinomial}(\boldsymbol{\phi}_{z})$. 
\end{enumerate}

The Biterm Topic Model (BTM) estimates the probability of observing word pairs (biterms) across the entire corpus, rather than within individual documents. The resulting joint likelihood over all biterms $
B$ is expressed as
\begin{equation}\label{eq:1}
   p(B) = \prod_{i,j}\sum_{z}\boldsymbol\theta_z \, \boldsymbol\phi_{i|z} \, \boldsymbol\phi_{j|z}.
\end{equation}
This formulation allows the model to capture global word co-occurrence patterns that help infer more stable topic–word associations. The conditional posterior for assigning a biterm to topic assignment $z$ is given by
\begin{equation}\label{eq:2}
    p(z|\mathbf{z}_{-b},\mathbf{B},\alpha,\beta) \propto (n_{z}+\alpha)\frac{(n_{w_i|z}+\beta)(n_{w_j|z}+\beta)}{(\sum_{w}{n_{w|z}}+M\beta)^2},
\end{equation}
where $n_z$ is the number of biterms currently assigned to topic $z$, $n_{w|z}$ is the number of times word $w$ is assigned to topic $z$, and $\mathbf{z}_{-b}$ denotes topic assignments excluding the current biterm. Intuitively, this expression increases the probability of assigning a biterm to topics that already have high counts for its component words, while $\alpha$ and $\beta$ serve as smoothing hyperparameters controlling topic and word diversity.

To address convergence issues that may arise from direct Gibbs sampling, we employ collapsed Gibbs sampling, which integrates out $\boldsymbol{\theta}$ and $\boldsymbol{\phi}_{z}$ by exploiting conjugacy of the Dirichlet–Multinomial distributions \citep{liu1994collapsed}. After sufficient sampling iterations, the posterior estimates of topic–word distributions and topic proportions are: 
\begin{equation}\label{eq:3}
    \boldsymbol\phi_{w|z} = \frac{n_{w|z}+\beta}{\sum_{w}{n_{w|z}}+M\beta}, \quad 
    \boldsymbol\theta_{z}=\frac{n_{z}+\alpha}{|B|+K\alpha}.
\end{equation} 
\vspace{-1em}

\begin{algorithm}
\caption{Collapsed Gibbs Sampler for BTM} 
\textbf{Input}: number of topics $K$, hyperparameters $\alpha$, $\beta$, biterm set $\mathbf{B}$  

\textbf{Output}: topic–word distributions $\boldsymbol\phi_{w|z}$ and topic proportions $\boldsymbol\theta_z$
	\begin{algorithmic}[1]
	\State Initialize topic assignments randomly for all biterms
		\For {iteration $=1,2,\ldots,N$}
			\For {each biterm $b \in \mathbf{B}$}
				\State Sample $z_{b}$ from $p(z|\mathbf{z}_{-b},B,\alpha,\beta)$ 
				\State Update counts $n_{w|z}$ and $n_z$ 
				\State Compute parameters $\boldsymbol\phi_{w|z}$ and $\boldsymbol\theta_z$
			\EndFor
		\EndFor
	\end{algorithmic} 
\end{algorithm}
\vspace{-1em}

Here, $M$ denotes the total number of unique words in the corpus and $K$ represents the number of latent topics. Accordingly, the terms $M\beta$ and $K\alpha$ act as normalization constants arising from the Dirichlet–Multinomial conjugacy, where $\beta$ and $\alpha$ control the concentration of word and topic distributions, respectively. Larger values of $\beta$ or $\alpha$ lead to smoother, less peaked distributions, while smaller values encourage more concentrated topic–word associations.

Specifically, we construct a word–topic probability matrix $\mathbf{X}$ of dimension $N \times P$, where $N$ is the number of words and $P$ is the number of topics. To identify characteristic words, we compute two measures for each row of $\mathbf{X}$: (i) the coefficient of variation across topics, and (ii) the maximum probability. The coefficient of variation captures the relative dispersion of a word’s probabilities across topics, while the maximum probability identifies whether the word is strongly associated with at least one topic. Words with both high dispersion and high maximum probability are retained as representative terms, enabling more interpretable characterization of latent topics.

\subsection{Clustering Topics via Latent Space Item Response Model}

We estimate interactions among topics and visualize their relationships by embedding them into a latent interaction map. 
To achieve this, we follow the approach of \citet{jeon2025network}, who applied LSIRM to topic–word distributions, and employ its Gaussian version in our setting. Specifically, we use the Gaussian LSIRM to represent the bipartite structure between topics and words, where topics are regarded as ``items'' and words as ``respondents'' as below: 
\begin{equation}
    x_{i,p} \mid \boldsymbol{\Theta} = \boldsymbol{a}_i +\boldsymbol{b}_p - || {\bf v}_i - {\bf u}_p || + \epsilon_{i,p}, \\ 
    \quad \epsilon_{i,p} \sim \mbox{N} (0, \sigma^2),
\end{equation}
where $x_{i,p}$ denotes the probability of word $i$ belonging to topic $p$. The parameters $\boldsymbol{a}_i$ and $\boldsymbol{b}_p$ represent the intrinsic activity of word 
$i$ (i.e., how generally a word tends to appear across topics) and the popularity or overall strength of topic $p$, respectively.  The Euclidean distance  $||{\bf v}_i - {\bf u}_p||$ measures how semantically close the word and topic are in the latent space, where smaller distances indicate stronger associations. Bayesian inference is employed to estimate the full parameter set $\boldsymbol{\Theta}=\{\boldsymbol{a}, \boldsymbol{b}, \mathbf{U}, \mathbf{V}\}$ with appropriate priors, and parameters are sampled using Markov chain Monte Carlo (MCMC). As a result, we obtain latent positions of topics $\mathbf{v}_i$ in $\mathbb{R}^d$, forming the topic coordinate matrix ${\bf A} \in \mathbb{R}^{d \times P}$.

Once the latent positions of topics are estimated, we proceed to cluster the topics in order to identify groups with similar semantic characteristics. To this end, we apply the K-means clustering algorithm to the latent position matrix $\mathbf{A}$. The K-means method partitions the $P$ topics into $C$ disjoint clusters $\{C_1, \ldots, C_C\}$ by minimizing the within-cluster sum of squared distances:
\begin{equation}
    \underset{C_1,\ldots,C_C}{\arg\min} \sum_{c=1}^{C} \sum_{\mathbf{v}_i \in C_c} \left\lVert \mathbf{v}_i - \boldsymbol{\mu}_c \right\rVert^2,
\end{equation}
where $\boldsymbol{\mu}_c$ denotes the centroid of cluster $C_c$. This clustering step groups topics that are located close together in the latent space, thereby reflecting their semantic similarity as derived from the topic–word distributions. 

The resulting framework enables us not only to visualize relationships among topics through their latent embeddings, but also to categorize them into interpretable clusters. Specifically, LSIRM provides a probabilistic mechanism to embed topics in a common latent space, and K-means clustering on the estimated positions further organizes these topics into coherent groups. This joint approach allows us to explore both the global structure (via the interaction map) and the local grouping (via cluster assignments) of topics inferred from the text data.

\subsection{Risk Index Factor via Lasso Regression}

To assess the contribution of words in explaining fire-related financial losses, we model the expected property damage amount $y$ for each incident report as the outcome variable and the extracted words as predictors $x$. Specifically, let $\mathbf{y} = (y_1, \ldots, y_n)$ denote the vector of estimated damages across $n$ reports, and let $\mathbf{X} \in \mathbb{R}^{n \times p}$ be the document–term matrix, where each column corresponds to one of the $p$ words. We fit a Lasso regression model~\citep{tibshirani1996regression} of the form
\begin{equation}
    \hat{\boldsymbol{\beta}} = \arg\min_{\boldsymbol{\beta}} \Bigg\{ \frac{1}{2n}\lVert \mathbf{y} - \mathbf{X}\boldsymbol{\beta} \rVert_2^2 + \lambda \lVert \boldsymbol{\beta} \rVert_1 \Bigg\},
\end{equation}
where $\lambda > 0$ is a tuning parameter. The $\ell_1$ penalty induces sparsity in the estimated coefficients, effectively performing variable selection by shrinking many coefficients toward zero. This property enables us to identify only those words that exhibit substantial predictive power for property damage amounts.  

The estimated coefficients $\hat{\beta}_j$ quantify the influence of word $j$ on expected fire-related losses. Coefficients close to zero indicate little or no association between the word and the damage amount, while coefficients with larger magnitude capture stronger effects. Moreover, the sign of $\hat{\beta}_j$ allows direct interpretation: negative values imply that the presence of a word is associated with lower property damage, whereas positive values suggest that the word signals higher expected losses. In this way, the Lasso framework provides a principled mechanism to evaluate and interpret the importance of words in the context of fire incident reports. Building on the estimated coefficients, we further construct a Risk Index to quantify word-level contributions in a structured manner. Specifically, we derive the index along three complementary dimensions.

First, within each cluster of semantically related words, the estimated coefficients are rescaled using a min–max transformation to lie between 0 and 1. This yields the \textit{i) Risk Index within Cluster} ($\boldsymbol{\gamma}_{i,c}, i=1,\cdots,n_c, c = 1,\cdots, C$), which allows for comparison of words relative to others in the same group. Second, to assess risk at the cluster level, we compute the mean coefficient value across words in each cluster $c$ and again apply min–max scaling to obtain a \textit{ii) Cluster-Level Risk Index} between 0 and 1 ($\boldsymbol{\delta}_{c}, c= 1,\cdots,C)$. This provides an interpretable measure of the overall risk associated with each cluster. Finally, the \textit{iii) Overall Risk Index ($\boldsymbol{\rho}_{i,c}$)} for a word is defined as the average of its within-cluster score and the risk score of its corresponding cluster. In this way, the framework jointly accounts for both the local importance of a word relative to its peers and the broader risk tendency of its semantic group. The resulting index serves as a principled indicator to identify key linguistic factors associated with higher fire-related property damages in incident reports.

\section{Results}

\subsection{Topic Characterization through Words}

As illustrated in Step 2 of Figure 1, the fire investigation texts were first analyzed through topic modeling to identify latent semantic structures. Table~\ref{tab:topic_names} presents the resulting thematic groups, providing an overview of how the documents are organized into distinct topics. In addition, Table~\ref{tab:topic_top5words} presents the distribution of words derived from the estimated topic–word probabilities ($\boldsymbol{\phi}_{w|z}$ obtained in the topic modeling stage. The table lists the top five terms with the highest posterior probabilities for each topic, highlighting representative keywords that characterize the semantic content of each domain.

\vspace{-1em}

\begin{table}[H]
\centering
\scriptsize
\caption{Mapping of extracted topics from fire investigation records to their descriptive names. The thematic categories are assigned based on high-probability keywords (summarized in Table~\ref{tab:topic_top5words}) and domain-specific interpretation.}
\label{tab:topic_names}
\begin{adjustbox}{max width=\linewidth}
\begin{tabular}{|c|p{15cm}|}
\hline
\textbf{Topic} & \textbf{Name} \\
\hline
Topic 1 & Flammable vapor ignition due to the use of organic solvents such as cleaning solutions \\
Topic 2 & Explosions caused by sparks generated from friction or static electricity due to the blending of flammable or combustible raw materials \\
Topic 3 & Fires caused by common electrical factors \\
Topic 4 & Ignition caused by sludge accumulation within ventilation equipment \\
Topic 5 & Ignition caused by electrical heating \\
Topic 6 & Related to dust collection equipment \\
Topic 7 & Spontaneous combustion caused by improper storage or containment of flammable residues \\
Topic 8 & Fire caused by forklift operation \\
Topic 9 & Chemical explosion occurring in the reactor \\
Topic 10 & Ignition of residue accumulated in ducts connected to or adjacent to dust collection equipment \\
Topic 11 & Due to improper use of drying equipment \\
Topic 12 & Fire caused by improper use of a banbury mixer for rubber molding \\
Topic 13 & Combustible materials ignited due to welding spark during hot work \\
Topic 14 & Fire caused by ignition sources in machinery with operation motor, such as air compressors \\
Topic 15 & Ignition of waste materials due to improper disposal of cigarette butts \\
\hline
\end{tabular}
\end{adjustbox}
\end{table}
\vspace{-0.7em}

Topic 1 reflects incidents associated with oil vapors generated from organic solvents or related chemical processes, frequently occurring in cleaning operations or wastewater treatment facilities. Topic 2 captures fire risks arising from the ignition of combustible materials due to friction or static electricity during equipment operation, as suggested by keywords such as \emph{mixer} and \emph{drum can}. Topic 3 represents general electrical fires, characterized by terms such as \emph{electrical short circuit} and \emph{circuit breaker}, while Topic 5 is more narrowly related to electrical heating sources (e.g., \emph{air conditioners} and \emph{heating wires}). Topic 4 emphasizes ignition triggered by sludge accumulation within ventilation systems, particularly in laboratory environments.

Topic 6 highlights fires directly linked to dust collection equipment, where flames may propagate through ducts or filter systems containing combustible dust. In contrast, Topic 10, though conceptually related, places less emphasis on dust collection equipment itself and instead indicates fire hazards involving adjacent facilities, such as \emph{ventilation ducts}, \emph{plastics}, and \emph{drying equipment}. Topic 7 is dominated by spontaneous combustion events caused by the improper storage of combustible residues, including processed byproducts such as \emph{sesame dregs}. Topic 8 involves forklift-related fires, which occur across both chemical and general factory settings, often due to battery or engine compartment failures. Topic 9 concerns reactor-related incidents in chemical plants, where abnormal reactions generate oil vapors leading to explosions.

Other topics describe equipment-specific or context-specific fire causes. Topic 11 focuses on drying equipment, particularly in cases involving powders such as \emph{silicon}. Topic 12 reflects fires ignited by the thermal oil of \emph{banbury} mixing equipment. Topics 13 through 15 capture more general factory-related accidents: Topic 13 refers to welding-induced ignition near cooling towers or sandwich panels; Topic 14 highlights motor-related electrical fires in air compressors; and Topic 15 illustrates landfill or waste-area fires, frequently initiated by discarded cigarette butts.

While biterm modeling provides an effective means of partitioning documents into topics, it assigns every word to all topics, which limits its ability to capture direct relationships among words. Since the primary goal of this study is to identify the words that carry substantial meaning within fire investigation documents, it is essential to explore how words are grouped through their interactions. In this respect, topic–word distributions offer probabilistic information on the degree of association between words and topics, thereby serving as a valuable resource for indirectly inferring inter-word relationships. Leveraging this information enables us to examine documents not only at the level of topics, but also from the perspective of word-level associations, ultimately facilitating a more interpretable summarization of fire-related narratives.

\vspace{-0.5em}
\begin{table}[H]
\centering
\scriptsize
\caption{Top five representative words for each extracted topic based on topic word probability distributions. These high probability words provide the basis for interpreting the thematic characteristics of the topics.}
\label{tab:topic_top5words}
\begin{adjustbox}{max width=\linewidth}
\begin{tabular}{|c|p{13cm}|}
\hline
\textbf{Topic} & \textbf{Top 5 Words} \\
\hline
Topic 1 & cleaning, cleaning room, wastewater treatment plant, agitator, machine room \\
Topic 2 & mixer, drum can, pallet, plastic, explosion \\
Topic 3 & electrical short circuit, electrical distribution, flame, circuit breaker, SWGR \\
Topic 4 & extractor hood, oven, laboratory, transformer, electricity \\
Topic 5 & heating wire, outdoor unit of air conditioner, thermal/acoustical insulation, air conditioner, rooftop \\
Topic 6 & dust collection equipment, duct, flame, plenty of, filter system \\
Topic 7 & storage, sesame dregs, residues, spontaneous combustion, storage room \\
Topic 8 & forklift, battery, electrical short circuit, distribution, engine compartment \\
Topic 9 & explosion, container, reactor, oil vapor, static electricity \\
Topic 10 & dust collection equipment, duct, flame, ventilation duct, plastic \\
Topic 11 & drying equipment, powder, silicon, base material, storage \\
Topic 12 & base material, mixing equipment, Banbury, thermal oil, flame \\
Topic 13 & welding, cooling tower, flame, sandwich panel, acetylene \\
Topic 14 & air compressor, motor, electrical short circuit, vulcanizer, solenoid valve \\
Topic 15 & waste materials, cigarette butts, waste, stacking, flame \\
\hline
\end{tabular}
\end{adjustbox}
\end{table}

\subsection{Thematic Aggregation of Topics}

\begin{figure}[ht]
\centering\includegraphics[scale=0.55]{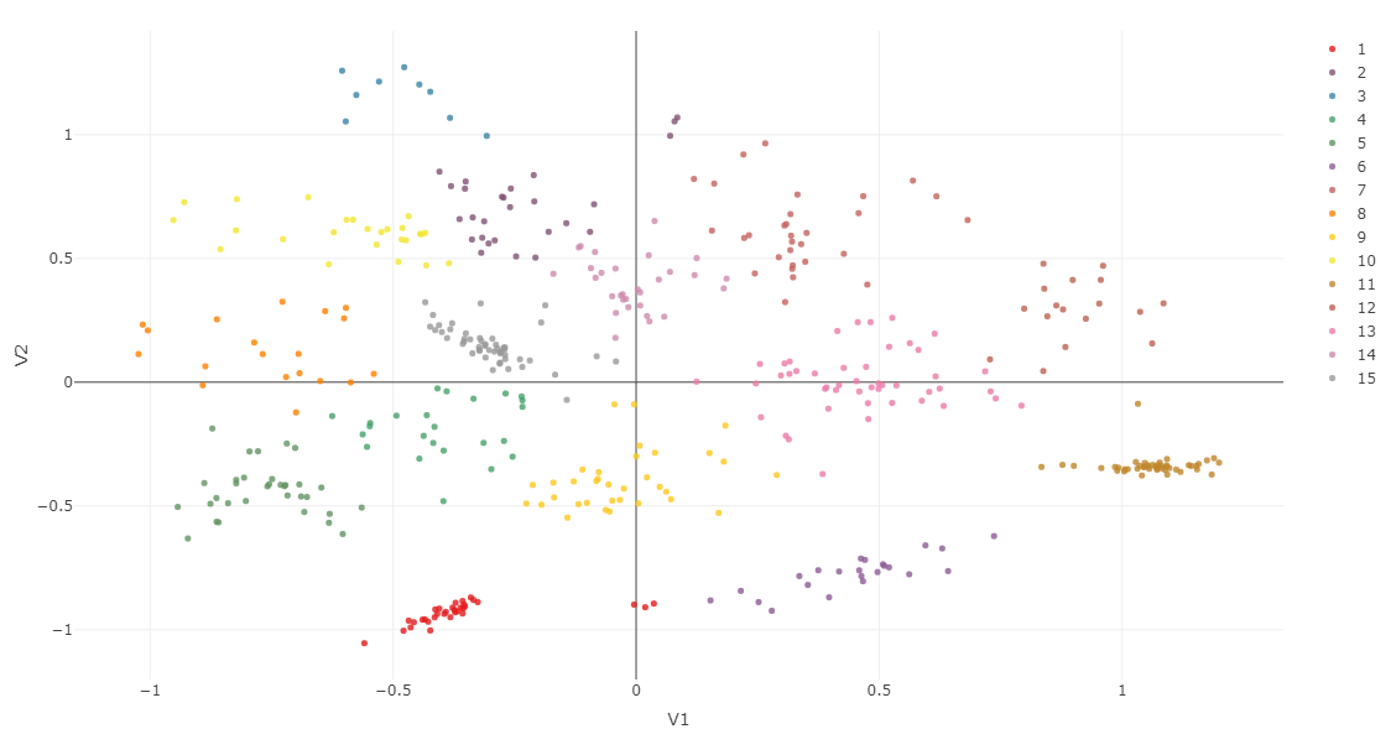}
\caption{Two-dimensional latent representation of words with clustering results obtained using the $k$-means algorithm. Distinct colors indicate the 15 identified clusters, showing how semantically related words are grouped in close proximity.}
\label{fig:topics2}
\end{figure}

Building upon the topic–word distributions, we further inferred interactions among words to explore higher-level thematic structures. Figure~\ref{fig:topics2} presents the two-dimensional projection of the latent positions of the 449 words introduced in Section~2.1, with clusters identified using the $k$-means algorithm. The visualization highlights clusters by distinct colors, showing how semantically related words are grouped in close proximity. Based on model selection criteria, a total of 15 clusters were identified as valid. As shown in the Figure~\ref{fig:topics2}, words located near one another in the latent space tend to form coherent clusters, thereby capturing meaningful associations beyond the topic-level representation.

The interpretation of clusters requires careful consideration because the meaning of a single word is better understood in relation to its neighboring terms. A word located at the center of a cluster typically exhibits a high probability of contributing to the generation of a topic, often alongside other words in the same group. However, the semantic similarity within a cluster may arise either from the joint contribution of the words themselves or from their shared functional context with alternative terms. Thus, rather than examining individual words in isolation, it is crucial to evaluate their surrounding vocabulary to delineate the collective meaning of each cluster. In this sense, clusters serve as categorical units in which semantic coherence emerges from localized word proximities, allowing for the clarification of latent thematic structures.

Following Step 3 in Figure 1, the estimated latent positions of words were used to cluster semantically related terms based on their pairwise distances. Table~\ref{tab:cluster_words} presents the top 10 representative keywords for each cluster, arranged according to their proximity to other words within the same cluster. This ordering enables a more precise characterization of the distinctive features of each cluster. While the identified clusters are not directly labeled as specific fire causes, their semantic coherence often reflects shared causal contexts. Accordingly, these clusters can be interpreted as data-driven representations of latent cause structures underlying fire incidents. Cluster 1 is anchored by the central term \emph{distillation column}, which suggests the risk of explosion due to chemical leaks or failures in pressure control. Surrounding words such as \emph{exposure}, \emph{solvent}, \emph{heptane}, and \emph{Silanes manufacturing} reinforce the theme of hazardous material leakage leading to fires or explosions. Accordingly, Cluster 1 is labeled as \emph{Fire or explosion caused by hazardous material leakage}.

Cluster 2 excludes the generic fire-related term \emph{ignition source} and instead centers on \emph{polishing}, which can produce ignition hazards when frictional heat contacts combustible materials. Neighboring words such as \emph{painting}, \emph{floor}, \emph{waste}, and \emph{interior wall} indicate that this cluster corresponds to \emph{Ignition from heat accumulation near combustible interior materials}.

Cluster 3 is defined by the term \emph{incinerate}, reflecting the inherent fire risks of incineration processes. Associated words such as \emph{waste wood}, \emph{interior materials}, \emph{heat}, and \emph{urethane} highlight ignition due to residual heat, while \emph{cutting machines} and \emph{dust collection equipment} suggest mechanical sources of smoldering. Thus, Cluster 3 is interpreted as \emph{Ignition due to residual heat post-work with combustibles}.

Cluster 4 revolves around the term \emph{heat of reaction}, but further interpretation requires contextualization with neighboring terms such as \emph{expired reagents}, \emph{corn}, and \emph{cooking oil}. These materials, when awaiting disposal, pose risks of spontaneous combustion, particularly when combined with \emph{oxygen}, \emph{heat waves}, or \emph{rainwater}. Hence, this cluster represents \emph{Spontaneous combustion from abandoned chemicals and oils}.

Cluster 5 is structured around \emph{abnormal reaction}, which links to numerous other terms and signals fires or explosions during abnormal chemical processing. Words such as \emph{pharmaceutical}, \emph{film}, \emph{oil}, and \emph{epoxy} point toward industrial settings where chemical instability can result in severe accidents. This cluster is labeled as \emph{Fires or explosions from abnormal reactions during chemical handling}.

Cluster 6 includes terms like \emph{storage}, \emph{cable}, \emph{control box}, and \emph{electrical circuit board}, reflecting general electrical fires not tied to specific chemical processes. Its theme is summarized as \emph{Electrical fires in areas handling flammable materials}.

Cluster 7 is characterized by \emph{coating}, which signals risks associated with flammable paints. Surrounding terms such as \emph{corrosion}, \emph{cutting oil}, \emph{grinders}, and \emph{presses} suggest ignition by sparks or friction from industrial machinery. Accordingly, Cluster 7 is described as \emph{Ignition of flammable substances (e.g., machine oil) by friction heat or spark from machinery}.

Cluster 8 is centered on \emph{vapor}, a strong indicator of fire hazards in chemical plants. Terms such as \emph{nucleic acid}, \emph{grease}, and \emph{hazardous materials} highlight risks arising from the ignition of volatile organic vapors, supporting the interpretation of \emph{Ignition of flammable vapor during organic solvent use}.

Cluster 9 highlights the term \emph{drying room}, referring to environments with elevated fire risks due to sustained high temperatures. Neighboring words, including \emph{cosmetic}, \emph{ventilation fans}, \emph{rotating}, \emph{small amount}, and \emph{decompose}, suggest fire hazards associated with oil vapor and production processes. Thus, Cluster 9 is classified as \emph{Ignition from temperature rise in oil vapor areas}.

Cluster 13 incorporates \emph{dilution}, \emph{power outage}, \emph{tracking}, \emph{muller}, \emph{vulcanizer}, and \emph{melting fusion}, suggesting interactions between hazardous materials, electrical ignition sources, and mechanical heat. This combination indicates \emph{Ignition in areas with accumulated combustible dust and oil vapors}.

Cluster 14 includes terms such as \emph{micro}, \emph{seat pad}, \emph{deodorizing tower}, \emph{absorbent pad}, and \emph{gunnysack}, which are not individually definitive. However, the presence of \emph{compost} and \emph{sesame dregs} highlights substances prone to spontaneous combustion through oxidation heat. Accordingly, this cluster represents \emph{Spontaneous combustion from improper oil residues storage or disposal}.

Finally, Clusters 10, 11, 12, and 15 are each characterized by distinct keywords: Cluster 10 corresponds to \emph{Fire from ignition in accumulated combustibles}, Cluster 11 to \emph{Fires from electrical factors in process equipment sites}, Cluster 12 to \emph{Fires during machinery maintenance}, and Cluster 15 to \emph{Ignition from sparks inside dust collection equipment with filters and debris}. Importantly, these categories represent fire scenarios that extend beyond chemical plants to general factory environments.

\vspace{1em}

\begin{table}[H]
\centering
\scriptsize
\caption{Clusters of 10 representative keywords derived from interword correlations based on estimated latent positions.}
\label{tab:cluster_words}
\begin{adjustbox}{max width=\linewidth}
\begin{tabular}{|c|p{15cm}|}
\hline
\textbf{Cluster} & \textbf{10 Words} \\
\hline
1  & distillation column, exposure, solvent, photoinitiator, metal, cleaning solvent, heptane, nozzle, flexible, silanes manufacturing \\
2  & ignition source, polishing, painting, floor, laboratory, electricity, plenty of, scrap paper, waste, recycling \\
3  & incinerate, waste wood, interior material, heat, urethane, smoldering ignition, cutting machine, dust collection equipment, rubber, manufacturing machine \\
4  & heat of reaction, expired reagent, corn, cooking oil, storage tank, reagent, oxygen, eruption, heat wave, rainwater \\
5  & abnormal reaction, pharmaceutical, film, sheath heater, upper, suction, oil, wire mesh, epoxy, dust explosion \\
6  & storage, cable, malfunction, cool down, oil tank, control box, coil, electrical circuit board, refrigerator, fertilizer \\
7  & coating, injection, corrosion, cutting oil, grinder, press, repair, acetylene, scrubber, steam equipment \\
8  & vapor, nucleic acid, grease, leak, hazardous material, mix, paint, high temperature, large scale, thermal cutting \\
9  & drying room, cosmetic, pallet, ventilation fan, pressure, ceiling, rotation, small amount, decompose, lid \\
10 & preheat, ignition, stacked, accumulation, tire, moisture, activated carbon, semi-finished product, waste storage, welding \\
11 & cooling fan, service line, shut out, lower terminal, insulating oil, underground, fan belt, switch, electric current, analysis lab \\
12 & closing, charging equipment, flammability, maintenance, automation, repair, cooling machine, machine room, gas torch, arc \\
13 & dilution, power outage, cover, immediate upper, tracking, prefabricated, vacuum, muller, vulcanizer, melting fusion \\
14 & micro, seat pad, deodorizing tower, absorbent pad, gunnysack, ton bag, compost, boiler room, sesame dregs, oxidation heat \\
15 & belt, particle, capture, aluminum, inlet, rubber department, wood chip, blowing, metal powder, abradant \\
\hline
\end{tabular}
\end{adjustbox}
\end{table}
\vspace{1em}

\begin{figure}[ht]
\centering\includegraphics[scale=0.45]{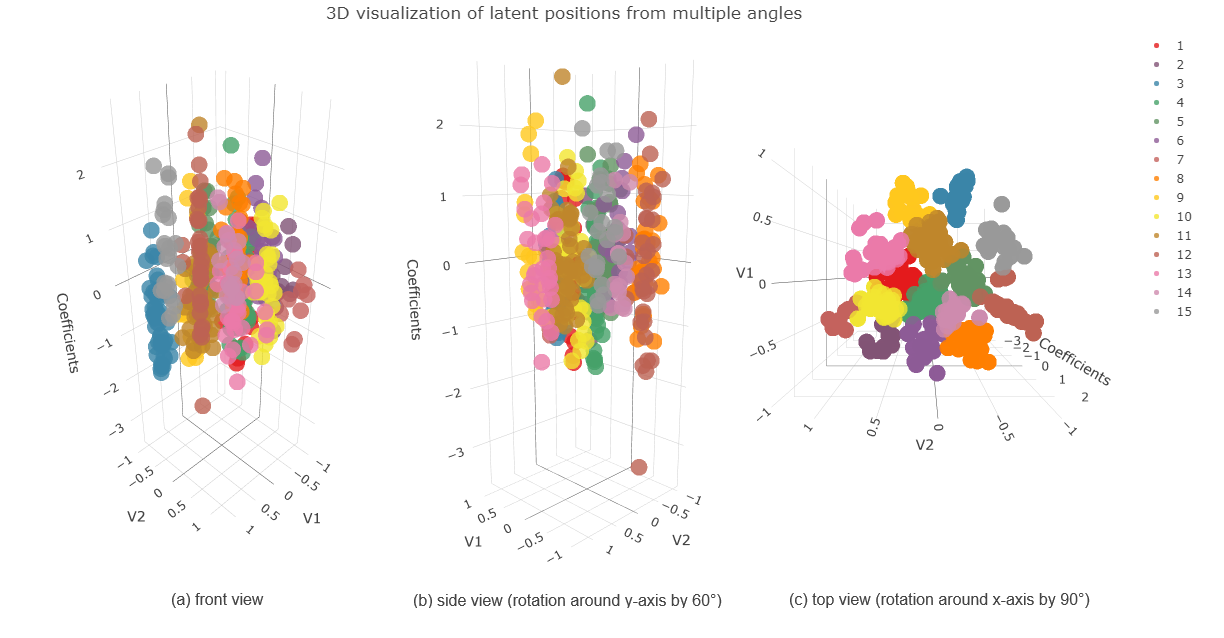}
\caption{Three-dimensional visualization of latent word embeddings from multiple perspectives. The $x-$ and $y-$axes represent latent positions, while the $z-$axis corresponds to coefficients estimated from the LASSO regression on fire-damage insurance claims. Panels (a), (b), and (c) show the front, side, and top views, respectively. Colors indicate clusters identified through k-means grouping.}
\label{fig:coefficients}
\end{figure}

Taken together, the interpretation of clusters provides meaningful categorical insights into the semantic structure of fire investigation records. However, semantic similarity alone does not capture the extent of economic severity associated with each term. To address this, we incorporated fire-damage estimates  into the analysis by estimating word-level coefficients via LASSO regression. This approach allows us to link clusters with the magnitude of potential financial losses. The coefficients estimated for each word quantify its relative contribution to explaining variations in property damage estimates. Figure~\ref{fig:coefficients} presents a three-dimensional visualization, where the horizontal axes represent the latent positions of words and the vertical axis corresponds to their regression coefficients. Each colored dot denotes a single word, and the color indicates its cluster membership consistent with the grouping shown in Figure \ref{fig:topics2}. This visualization allows direct comparison between semantic proximity (latent positions) and the estimated contribution of each word to financial damage. Words located at higher positions along the coefficient axis correspond to terms associated with higher risk of financial loss, whereas clusters with generally lower coefficient values indicate groups of words linked to less severe incidents. By examining the clusters from multiple viewing angles, one can observe that several clusters exhibit a wide range of coefficient magnitudes, suggesting heterogeneous risk levels within the same semantic domain, while others maintain consistently low coefficients, reflecting more homogeneous, low-risk contexts.

As illustrated, even within the same semantic cluster, words exhibit heterogeneous patterns: some words have positive coefficients, indicating stronger associations with larger property losses, while others display negative coefficients, reflecting lower associated damage levels. This observation highlights that clusters capture thematic similarity but do not necessarily imply uniform economic consequences. Hence, incorporating these regression coefficients into subsequent analyses provides an additional, economically grounded perspective. In particular, the integration of semantic clustering with property damage-based coefficients motivates the construction of a risk index that simultaneously reflects linguistic structure and financial severity, thereby offering a more comprehensive measure of fire-related risks.

\subsection{Risk Index Estimation}

The preceding analyses demonstrate that clusters derived from topic–word distributions provide semantically coherent categories of fire-related terms, while regression coefficients estimated from property damage amounts capture the associated economic severity. Importantly, these two perspectives highlight complementary aspects of risk: semantic clusters reflect the contextual mechanisms of fire occurrence, whereas coefficients quantify their financial impact. Moreover, as shown in the regression analysis, even words within the same cluster may exhibit heterogeneous patterns of association with damage amounts, indicating that semantic similarity alone is insufficient for fully characterizing fire risk.

To address this limitation, we propose the construction of a composite risk index that integrates both linguistic and economic dimensions. To quantify the relative contribution of words and clusters to fire-related incidents, we define three levels of risk indices: the word-level index $\gamma_{i,c}$, the cluster-level index $\delta_{c}$, and the overall word index $\rho_{i,c}$. Each measure captures a distinct dimension of risk, ranging from fine-grained lexical associations to broader thematic categories. By jointly considering (i) cluster-level semantic associations and (ii) word-level coefficients derived from loss data, the risk index provides a systematic measure of fire risk that is interpretable in terms of language use and grounded in economic outcomes.

\paragraph{Word-level Risk Index ($\gamma_{i,c}$).}

As described in Step 4 of Figure 1, the Lasso-based risk modeling was applied to estimate the $\gamma_{i,c}$ that measures the relative contribution of word $i$ within cluster $c$ to the estimation of property damage amounts. A higher value indicates that the word is more strongly associated with larger expected losses compared to other words in the same cluster. Table~\ref{tab:cluster_rif} presents the top ten words with the highest $\gamma_{i,c}$ values in each cluster, along with their Risk Index. For example, in Cluster~2 (\emph{Ignition from heat accumulation near combustible interior materials}) and Cluster~7 (\emph{Ignition of flammable substances by friction heat or spark from machinery}), words such as \emph{flame}, \emph{heat}, and \emph{grinder} show high $\gamma_{i,c}$ values, reflecting their strong linkage with higher levels of estimated property damage. Thus, $\gamma_{i,c}$ highlights words whose relative importance provides insight into the financial risk implications captured within each cluster.

\begin{landscape}
\begin{table}[htbp]
\centering
\scriptsize
\begin{adjustbox}{max width=\linewidth}
\begin{tabular}{|c|p{2.3cm}|p{2.3cm}|p{2.3cm}|p{2.3cm}|p{2.3cm}|p{2.3cm}|p{2.3cm}|p{2.3cm}|p{2.3cm}|p{2.3cm}|}
\hline
\textbf{Cluster} & \textbf{Word 1} & \textbf{Word 2} & \textbf{Word 3} & \textbf{Word 4} & \textbf{Word 5} & \textbf{Word 6} & \textbf{Word 7} & \textbf{Word 8} & \textbf{Word 9} & \textbf{Word 10} \\
\hline
1  & photoinitiator (1.000) & ethyl acetate (0.911) & reactor (0.846) & heptane (0.814) & sclerotic (0.809) & centrifuge (0.798) & lesk (0.781) & cleaning solvent (0.741) & solvent (0.739) & chemical reaction (0.736) \\
2  & flame (1.000) & plenty of (0.937) & wall (0.893) & interior wall (0.891) & laboratory (0.858) & floor (0.856) & radiant heat (0.834) & laboratory (0.736) & insulator (0.712) & exterior wall (0.674) \\
3  & dust collection equipment (1.000) & urethane (0.907) & smoldering ignition (0.740) & rubber (0.714) & interior material (0.657) & heat (0.493) & cutting machine (0.445) & waste wood (0.082) & incinerate (0.000) & – (1.000) \\
4  & stacked goods (0.987) & oil vapor (0.970) & storage tank (0.675) & gas (0.599) & oxygen (0.581) & reagent (0.448) & dust collector (0.427) & waste (0.381) & dummy (0.317) & oil (0.296) \\
5  & manhole (1.000) & expander (0.949) & cover (0.939) & film (0.890) & epoxy (0.819) & mixer (0.801) & manufacturing building (0.798) & static electricity (0.683) & impurities (0.633) & physic (0.594) \\
6  & shipping area (1.000) & petroleum product (0.879) & condenser (0.878) & storage (0.745) & cable (0.708) & refrigerator (0.691) & arson (0.642) & crayon (0.539) & oil tank (0.484) & hazardous substances plant (0.461) \\
7  & grinder (1.000) & polystyrene (0.952) & fabric (0.909) & long time (0.902) & coating (0.816) & demolition (0.780) & manufacturing room (0.769) & panel (0.747) & electrical distribution (0.648) & cutting oil (0.637) \\
8  & chemical material (1.000) & hazardous material (0.852) & leak (0.753) & vapor (0.680) & metal plate (0.603) & base material (0.544) & nucleic acid (0.456) & spontaneous combustion (0.424) & grease (0.402) & large-scale (0.341) \\
9  & warehouse (1.000) & pallet (0.933) & storage (0.851) & cosmetic (0.825) & agitator (0.789) & drying room (0.712) & liquid (0.696) & secondary battery (0.693) & staff lounge (0.665) & heater rod (0.522) \\
10 & activated carbon (1.000) & welding (0.863) & remain (0.816) & moisture (0.588) & semi-finished product (0.572) & accumulation (0.395) & tire (0.393) & paint (0.378) & plastic (0.369) & ignition (0.367) \\
11 & service line (1.000) & packing room (0.869) & analysis lab (0.841) & interlayer short circuit (0.833) & distribution board (0.826) & molding machine (0.712) & unidentified short circuit (0.673) & power line (0.656) & trip (0.618) & air-compressor (0.616) \\
12 & automation (1.000) & flammable (0.938) & repair (0.843) & machine room (0.607) & insulation deterioration (0.541) & air-conditioner (0.356) & arc (0.328) & terminal (0.316) & battery (0.268) & outdoor unit of air-conditioner (0.258) \\
13 & vulcanizer (1.000) & prefabricated (0.662) & research building (0.553) & indoor wiring (0.500) & dilution (0.492) & complete product (0.482) & automobile parts (0.480) & filtering equipment (0.465) & boiler (0.460) & circuit breaker (0.455) \\
14 & roof (1.000) & entrance (0.969) & unidentified cause (0.889) & machine (0.885) & sandwich panel (0.829) & appliances (0.768) & dormitory (0.746) & compost (0.734) & boiler room (0.634) & deodorizing tower (0.609) \\
15 & lower (1.000) & explosion (0.769) & electrical equipment (0.684) & commissioning (0.622) & duct (0.621) & rigid polyurethane (0.605) & stacked materials (0.595) & melting equipment (0.574) & condensation (0.562) & freeze drier (0.548) \\
\hline
\end{tabular}
\end{adjustbox}
\caption{Cluster–Word table with Risk Index values. 
For each cluster, the top ten words with the highest word-level Risk Index ($\gamma_{i,c}$) are reported. 
These words represent the relatively high-scoring lexical elements within each cluster, indicating stronger contributions to the estimated property damage amounts.}
\label{tab:cluster_rif}
\end{table}
\end{landscape}

\paragraph{Cluster-level Risk Index ($\delta_{c}$).}
The index $\delta_{c}$ summarizes the risk associated with cluster $c$ as a whole, aggregating the estimated coefficients of its constituent words. A higher $\delta_{c}$ indicates that, on average, words belonging to this cluster are strongly predictive of higher property losses. Table~\ref{tab:cluster_rif_ranking} reports these values. The five clusters with the highest $\delta_{c}$ include: (i) \emph{Ignition of flammable vapor during organic solvent use}, (ii) \emph{Electrical fires in areas handling flammable materials}, (iii) \emph{Ignition from temperature rise in oil vapor areas}, (iv) \emph{Ignition from heat accumulation near combustible interior materials}, and (v) \emph{Fires or explosions from abnormal reactions during chemical handling}. These topics correspond to scenarios where ignition sources and flammable environments directly translate into severe financial consequences. By contrast, the clusters with the lowest $\delta_{c}$, such as \emph{Ignition from sparks inside dust collection equipment with filters and debris} or \emph{Fires from electrical factors in process equipment sites}, represent situations with relatively weaker association to large-scale losses. In this way, $\delta_{c}$ provides an interpretable measure of how strongly each cluster of fire-related factors contributes to financial risk.

\begin{table}[htbp]
\centering
\scriptsize
\begin{tabular}{|c|c|p{9cm}|}
\hline
\textbf{Cluster} & \textbf{Risk Index} & \textbf{Topic} \\
\hline
8  & 1.000 & Ignition of flammable vapor during organic solvent use \\
6  & 0.954 & Electrical fires in areas handling flammable materials \\
9  & 0.731 & Ignition from temperature rise in oil vapor areas \\
2  & 0.639 & Ignition from heat accumulation near combustible interior materials \\
5  & 0.626 & Fires or explosions from abnormal reactions during chemical handling \\
12 & 0.606 & Fires during machinery maintenance \\
7  & 0.561 & Ignition of flammable substances by friction heat or spark from machinery \\
4  & 0.534 & Spontaneous combustion from abandoned chemicals and oils \\
10 & 0.405 & Fire from ignition in accumulated combustibles \\
13 & 0.388 & Ignition in areas with accumulated combustible dust and oil vapors \\
14 & 0.344 & Spontaneous combustion from improper oil residues storage or disposal \\
3  & 0.294 & Ignition due to residual heat post-work with combustibles \\
1  & 0.216 & Fire or explosion caused by chemical leakage \\
15 & 0.205 & Ignition from sparks inside dust collection equipment with filters and debris \\
11 & 0.000 & Fires from electrical factors in process equipment sites \\
\hline
\end{tabular}
\caption{Cluster ranking by Risk Index values and their associated topics. 
The Risk Index ($\delta_{c}$) represents the aggregated risk contribution of each cluster, with higher values indicating strong fire-related financial losses.}
\label{tab:cluster_rif_ranking}
\end{table}

\paragraph{Overall Word Risk Index ($\rho_{i,c}$).}
The index $\rho_{i,c}$ extends beyond cluster membership to evaluate the global risk contribution of word $i$, accounting for its position across the latent embedding space. Table~\ref{tab:word_rif_cases} lists the top twenty words by $\rho_{i,c}$. For example, \emph{chemical material}, \emph{shipping area}, and \emph{hazardous material} emerge as the top three words. These terms directly connect to concrete accident scenarios such as the generation of flammable vapors from sludge leakage, electrical fires in distribution boards, and vapor leakage during hazardous material processing. The $\rho_{i,c}$ index therefore highlights words that carry not only lexical salience within clusters but also broader cross-cluster risk relevance. 

Collectively, these three indices provide a multi-layered framework: $\gamma_{i,c}$ identifies salient words within clusters, $\delta_{c}$ ranks the clusters by their aggregate hazard potential, and $\rho_{i,c}$ detects globally critical words linked to real-world accident narratives.

\begin{table}[htbp]
\centering
\scriptsize
\begin{adjustbox}{max width=\linewidth}
\begin{tabular}{|c|p{3cm}|p{2cm}|}
\hline
\textbf{No.} & \textbf{Word} & \textbf{Risk Factor} \\
\hline
1  & chemical material & 1.000 \\
2  & shipping area & 0.977 \\
3  & hazardous material & 0.926 \\
4  & petroleum product & 0.917 \\
5  & condenser & 0.916 \\
6  & warehouse & 0.866 \\
7  & storage facility & 0.850 \\
8  & pallet & 0.832 \\
9  & cable & 0.831 \\
10 & flame & 0.820 \\
11 & manhole & 0.813 \\
12 & automation & 0.803 \\
13 & storage & 0.791 \\
14 & plenty of & 0.788 \\
15 & expander & 0.788 \\
16 & cosmetic & 0.778 \\
17 & stacked goods & 0.761 \\
18 & agitator & 0.760 \\
19 & film & 0.758 \\
20 & spontaneous combustion & 0.712 \\
\hline
\end{tabular}
\end{adjustbox}
\caption{Word importance based on overall Risk Index ($\rho_{i,c}$). 
The table reports the top 20 words with the highest $\rho_{i,c}$ values, highlighting globally influential keywords strongly linked to insurance loss outcomes.}
\label{tab:word_rif_cases}
\end{table}

\section{Discussion}

\citet{wehmeier2016fire} analyzed the causes of fires in chemical plants based on incidents at a chemical–pharmaceutical company and categorized them as follows: Self-ignition (22\%), Hot running of moving parts (17\%), Welding (15\%), Electrostatic (14\%), Drying (10\%), Repair/Maintenance (8\%), Leakage (7\%), and Electric (short-circuit) (7\%). However, he also concluded that fire investigations in the German chemical industry reveal complex and heterogeneous causal structures. Furthermore, because this analysis relied solely on accident-frequency statistics, it was insufficient for assessing \emph{risk} in the sense of loss severity. Complementarily, \citet{darbra2010domino} constructed a relative-probability event tree for major chemical accidents with domino effects, analyzing the causes, materials involved, effects and consequences, affected population, and the likelihood of specific accident sequences, thereby emphasizing the need to evaluate interconnected hazards rather than isolated causes.

To address these limitations, we replace frequency-based tallies with a composite, loss-aware scoring framework that links text-derived indicators to the scale of damage. The procedure is twofold: (i) estimate a risk score for each term that reflects its association with property-relevant losses, and (ii) interpret high-scoring terms by examining their local semantic neighborhoods to recover operational and environmental context.

\begin{figure}[ht]
\centering\includegraphics[scale=0.2]{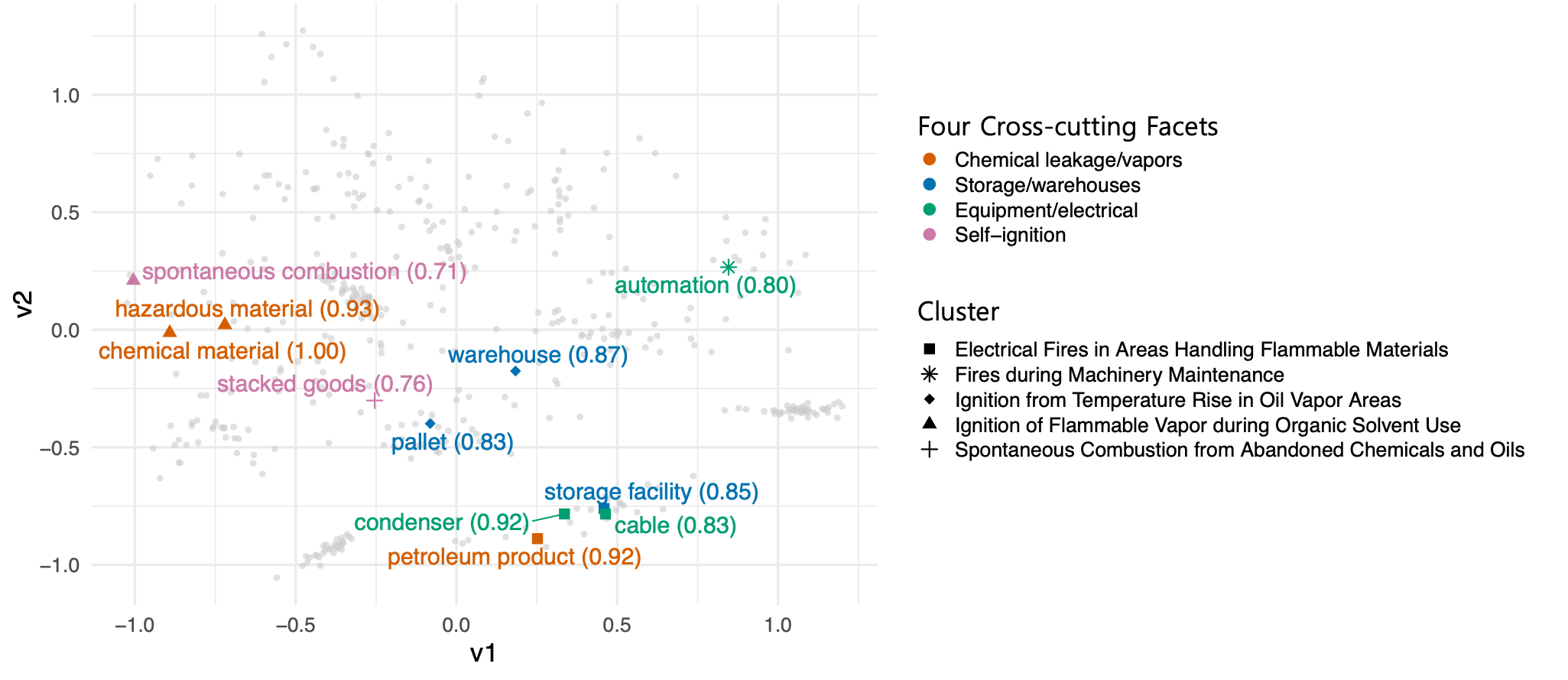}
\caption{Two-dimensional embedding (v1, v2) of fire-risk keywords, colored by four discussion facets and shaped by cluster assignments. Colored, labeled points are representative high-scoring terms (ranked by the Overall Word Risk Index $\rho_{i,c}$) in parentheses. Marker shapes encode cluster names as shown in the legend; gray points denote the remaining vocabulary.}
\label{fig:final}
\end{figure}

We instantiate this framework via the Overall Word Risk Index, $\rho_{i,c}$, which quantifies the association between terms and loss outcomes rather than raw frequency. Using $\rho_{i,c}$ to rank and organize the vocabulary in the embedding (Figure~\ref{fig:final}), we identify four cross-cutting facets of fire risk: (i) \emph{chemical leakage/vapors} (red), (ii) \emph{storage/warehouses} (blue), (iii) \emph{equipment/electrical} (green), and (iv) \emph{self-ignition} (pink). The latent coordinates in Figures~\ref{fig:topics2} and~\ref{fig:final} are derived from the same embedding space, while Figure~\ref{fig:final} highlights representative high-risk keywords and their corresponding topics to provide a more detailed interpretation across the four facets. In what follows, we analyze each facet and its representative cases.

\paragraph{Four Cross-cutting Facets of Fire Risk}

First, the highest-ranking terms such as chemical material, hazardous material, and petroleum product capture risks associated with chemical leakage and the generation of flammable vapors. These findings align with well-documented hazards in chemical industries, where improper handling, leakage, or inadequate containment of chemical substances often lead to large-scale fire and explosion events. The 2009 Jaipur crude oil pipeline tank fire in India is an example of how a fire originating from a hazardous material leak can spread. The accident resulted in 13 deaths and over 200 injuries, setting a record for the worst accident of its kind in India~\citep{lee2020study}. The representative cases illustrate how unintended chemical leaks can escalate into severe incidents with significant property and human losses. 

Second, structural and operational factors such as warehouses, storage facilities, and pallets are also prominent, underscoring the critical role of storage conditions and facility maintenance. Prior research on cold-chain logistics demonstrates that pallet stacking and package design strongly influence airflow and heat transfer within storage rooms, where insufficient ventilation can generate localized hotspots and elevate fire hazards~\citep{ahmad2022interactions}. In the context of energy storage, investigations highlight that inadequate cooling and airtight designs in portable energy storage systems promote overheating and exacerbate ignition risk~\citep{eslami2023review}. Similarly, studies on lithium-ion batteries reveal that poor thermal management and accumulation of flammable gases in confined storage spaces can accelerate thermal runaway and combustion~\citep{feng2018thermal}. Together, these findings indicate that storage practices, ranging from palletized materials to battery warehouses, interact with flammable substances in ways that compound ignition potential, while recurring references to ventilation-related equipment emphasize how failures in airflow management can amplify both the likelihood and severity of fire incidents.

Third, a notable set of keywords points to ignition sources related to malfunctions or inadequate maintenance of electrical and mechanical equipment, such as condenser, cable, automation. These emphasize that not only causes directly related to chemical processes, but also the integrity of auxiliary mechanical and electrical systems, play a crucial role in fire risk. Electrical fires are among the most universal causes of fire. Korean statistics show that 27.4\% of fires incidents from 1996 to 2021 were attributed to electrical factors~\citep{lee2023overview}, suggesting that the chemical industry is no exception to this risk. These findings suggest that the damage resulting from contact between common ignition sources and flammable raw materials or products handled in chemical plants can be significantly greater. The cases show that malfunctioning condensers, short-circuited cables, or maintenance using gas torches can trigger fires even in otherwise controlled environments.

Finally, terms such as spontaneous combustion and stacked goods reflect risks arising from self-ignition processes of chemicals and oil residues produced or derived in chemical plants, as well as environmental conditions. Recent studies on biomass storage have shown that moisture exchange, oxygen penetration, and heat accumulation can interact to induce self-heating and ultimately spontaneous ignition~\citep{wei2023modelling}. Similarly, investigations in mine waste dumps highlight that spontaneous combustion can compromise geomechanical stability, illustrating how natural or reactive processes extend the scope of fire hazards beyond purely mechanical or chemical failures~\citep{nguyen2025review}. These findings underscore the importance of incorporating self-ignition phenomena into fire risk assessments for chemical facilities and storage environments.

Taken together, these findings demonstrate that high-risk keywords are not confined to a single category but span across chemical substances, storage practices, equipment reliability, and environmental interactions. By transforming unstructured fire investigation narratives into analyzable data, our framework systematically extracted semantically related terms through latent topic and embedding models, contextualized them with representative fire cases, and linked them to property damage estimates. This integrative approach culminated in the development of a composite Risk Index, which quantifies the relative importance of text-derived indicators in relation to financial loss outcomes. The combination of statistical evidence with real-world loss information provides interpretability and practical relevance, showing how latent textual patterns can be operationalized into measurable safety metrics. In doing so, the proposed risk index offers a model-based tool for identifying actionable risk factors that bridge large-scale textual evidence with practical fire safety management.

\paragraph{Risk Index Perspective}
As summarized in Table~\ref{fig:final}, the \emph{chemical leakage/vapors} facet concentrates the highest $\rho_{i,c}$ values (e.g., \emph{chemical material}, \emph{hazardous material}, \emph{petroleum product}), indicating a stronger linkage to larger financial losses in our corpus—consistent with evidence that leakage and vapor formation are prominent drivers of severe incidents in chemical industries \citep{wehmeier2016fire,lee2020study}. By contrast, the \emph{self-ignition} facet (e.g., \emph{spontaneous combustion}, \emph{stacked goods}) exhibits lower $\rho_{i,c}$ values relative to the chemical-leakage/vapors facet in our corpus. This attenuation is consistent with three data-plausible mechanisms supported by prior work. First, self-heating often proceeds as a long-duration, low-temperature smouldering process with comparatively low heat-release rates and slow spread, which increases the opportunity for intervention before very large property losses accrue~\citep{rein2009smouldering,torero2020processes,santoso2019review}. Second, consistent with prior surveys of major incidents, very large losses in hydrocarbon processing frequently originate from sustained leaks that evolve into flash fires or vapour cloud explosions, rather than from long-duration smouldering scenarios \citep{atkinson2017vce,chang2006storagetank,darbra2010domino}. Third, contexts in which spontaneous combustion becomes catastrophic (e.g., biomass or mine-waste stockpiles) are emphasized in the engineering literature but are under-represented in our chemical special-building corpus, attenuating the empirical linkage between self-ignition terms and large losses \citep{wei2023modelling,nguyen2025review}. Interpreted this way, the relatively lower scores reflect dataset composition and event progression characteristics, rather than any contradiction with the established self-heating mechanism.

\paragraph{Why similarly clustered terms admit distinct operational readings.}

Although cluster membership (shapes in Fig.~\ref{fig:final}) reflects lexical neighborhoods, nearby terms can encode different operational contexts that co-occur in narratives. For example, \emph{storage} terms (blue) can sit beside \emph{equipment/electrical} terms (green) because pallet stacking and enclosure geometry restrict ventilation and heat rejection \citep{ahmad2022interactions}, thereby increasing the efficacy of routine electrical faults as ignition sources; likewise, \emph{self-ignition} terms (pink) may appear near \emph{leakage/vapor} terms (red) when oxygen ingress and heat accumulation in stacked goods elevate vapor formation and ignition potential \citep{wei2023modelling}. Conversely, process-centric terms (e.g., maintenance/automation) may be spatially offset when hot-work contexts (sparks, localized heating) are discussed apart from storage constraints, yet they remain semantically bridgeable whenever operations occur proximate to flammable inventories or confined airflow paths \citep{eslami2023pes,feng2018thermal}. Interpreting the embedding jointly with the loss-aware index \( \rho_{i,c} \) (Fig.~\ref{fig:final}) discriminates semantically adjacent yet economically distinct patterns: terms that lie close in the map but carry higher \( \rho_{i,c} \) mark contexts historically associated with larger property losses, whereas nearby low-\( \rho_{i,c} \) terms indicate operational exposure without the same tail severity. This joint reading proximity for mechanism, \( \rho_{i,c} \) for consequence which yields a ranked set of actionable priorities, directing inspection toward leakage/vapor configurations while maintaining vigilance for storage and self-ignition scenarios that can escalate under adverse conditions.

The proposed framework not only enables post-incident classification of fire causes but also provides a practical foundation for preventive fire management. By translating text-derived risk indicators into measurable patterns, the derived Risk Index helps identify high-risk operational contexts and materials before incidents occur. Table~\ref{tab:fire_prevention} illustrates how the ranked Risk Index can guide on-site inspections and safety checklists in industrial facilities. This approach supports decision-makers in prioritizing safety measures under limited manpower and budget, and establishes a systematic linkage between historical fire evidence and actionable preventive guidelines. Moreover, the framework can be extended by incorporating real-time sensor-based parameters such as temperature, dust concentration, or gas-leakage levels to enable proactive monitoring analogous to diagnostic systems in medicine. In practice, these parameters can be included as additional covariates in the Risk Index estimation stage, allowing the model to quantify how environmental conditions influence fire risk levels. By jointly modeling textual indicators and measurable physical factors, the extended framework can provide more timely and context-aware assessments, supporting preventive decision-making in high-risk industrial settings.

\begin{table}[htbp]
\centering
\footnotesize
\setlength{\tabcolsep}{4pt}   
\renewcommand{\arraystretch}{1.15} 
\caption{Example of Fire Prevention Guidelines for Industrial Facilities by Risk Index Ranking}
\label{tab:fire_prevention}
\begin{adjustbox}{max width=\linewidth} 
\begin{tabular}{|c|c|p{3.1cm}|p{3.0cm}|p{3.0cm}|p{6.5cm}|}
\hline
\textbf{Risk Index} & \textbf{Cluster No.} & \textbf{Topic} & \textbf{Main Keywords} & \textbf{Related Words} & \textbf{Fire and Explosion Prevention Checklist}\\
\hline
1.000 & 8 & Ignition of Flammable Vapor during Organic Solvent Use &
chemical material, hazardous material &
leak, paint, sludge, grease, vapor, mixing equipment &
\begin{minipage}[t]{\linewidth}\vspace{2pt}
\begin{itemize}[leftmargin=*, itemsep=1pt, topsep=1pt]
\item Are measures in place to prevent leakage of flammable chemical materials?
\item Is there any accumulation of sludge such as grease?
\item Is the ventilation system operating to prevent retention of flammable vapors?
\item Are vapors accumulating where sparks from mixing equipment could ignite?
\end{itemize}\vspace{2pt}
\end{minipage}\\
\hline
0.977 & 6 & Electrical Fires in Areas Handling Flammable Materials &
shipping area, distribution board & -- &
\begin{minipage}[t]{\linewidth}\vspace{2pt}
\begin{itemize}[leftmargin=*, itemsep=1pt, topsep=1pt]
\item Are wiring systems, including distribution boards in the shipping area, properly maintained?
\end{itemize}\vspace{2pt}
\end{minipage}\\
\hline
0.866 & 9 & Ignition from Temperature Rise in Oil Vapor Areas &
warehouse, breakdown, discharge, storage, secondary battery, synthetic resins, hot work equipment &
secondary battery, discharge, hot work equipment, synthetic resins &
\begin{minipage}[t]{\linewidth}\vspace{2pt}
\begin{itemize}[leftmargin=*, itemsep=1pt, topsep=1pt]
\item Are discharged secondary batteries stored safely?
\item Is hot work inside the warehouse performed under safe conditions?
\item Are combustible materials stored at a safe distance from hot work areas?
\end{itemize}\vspace{2pt}
\end{minipage}\\
\hline
\end{tabular}
\end{adjustbox}
\end{table}

\section{Conclusion}

This study presented a data-driven framework for quantifying and interpreting fire-related risk factors embedded in unstructured investigation narratives. By transforming textual records into latent topical structures and constructing an embedding-based network of word interactions, we identified semantically coherent clusters that capture operational and environmental contexts of fire incidents. The key contribution of this work lies in linking these text-derived semantic signals with structured financial loss data to develop a composite metric, the Overall Word Risk Index ($\rho_{i,c}$), which highlights words and topics most closely associated with real-world damage outcomes. The resulting framework provides an interpretable and loss-aware map of industrial fire risk, offering a foundation for evidence-based prevention and safety management.

In this study, the derived Risk Index can be considered useful as reference data for safety management and insurance underwriting of industrial sites. Within a workplace, the Risk Index can help assign priorities when establishing preventive measures against potential accidents inferred from topics and words. Given the limited manpower and budget typically available at industrial sites, allocating safety management resources in proportion to the degree of risk is of critical importance. In the context of insurance subscription, understanding the level of risk at a workplace is also of utmost priority. Although insurance premiums are generally determined based on statistical loss ratios by industrial sector, it remains difficult to assess the unique risks arising from a site’s specific operating conditions. In this regard, the Risk Index serves as a quantitative measure of a facility’s risk level and may provide valuable guidance for insurance underwriters when determining whether to underwrite facilities that involve substances or processes associated with higher risk scores.

Although the results are promising, potential threats to validity should also be acknowledged. The narratives used in this study were prepared by professional fire officers based on verified evidence and standardized terminology, ensuring a high degree of internal validity with minimal subjective bias. While supervised validation would ideally complement our approach, the current unsupervised framework incorporates expert interpretation to maintain analytical reliability. Regarding external validity, the proposed methodology could be extended to other domains such as manufacturing, logistics, construction, or residential and public facilities to assess its generalizability and adaptability across diverse fire contexts.

Future research will extend the current framework in two directions. First, we plan to develop methods for systematically tracking the temporal trends of risk factors, enabling the identification of how the prevalence and severity of specific hazards evolve over time. Such analyses will provide an evidence-based foundation for monitoring emerging risks and for designing timely interventions. Second, we aim to advance causal inference approaches tailored to text-derived risk indicators, with a particular focus on risky keywords identified in investigation records. By establishing causal relationships rather than mere associations, this line of work will enhance the interpretability of the extracted factors and strengthen their utility for policy and decision-making in fire risk management.

\bibliographystyle{plainnat}
\bibliography{reference}

\vspace{6pt} 

\end{document}